\documentclass[usenatbib]{mn2e}

\usepackage{graphicx}
\usepackage{amsmath}
\usepackage{url}

\usepackage{rccol}
\rcDecimalSign{.}

\def\aj{AJ}%
\def\apj{ApJ}%
\def\apjl{ApJ}%
\def\apjs{ApJS}%
\def\aap{A\&A}%
\def\mnras{MNRAS}%
\def\pasj{PASJ}%
\newcommand{\cpar}[1]{\left( #1 \right)}
\newcommand{\spar}[1]{\left[ #1 \right]}

\title[Gamma-ray variability of radio-loud narrow-line Seyfert 1
  galaxies]{Gamma-ray variability of radio-loud narrow-line Seyfert 1
  galaxies}

\author[G. Calderone et al.]
  {G. Calderone$^{1}$\thanks{E-mail: {\tt giorgio.calderone@mib.infn.it}},
  L. Foschini$^{2}$,
  G. Ghisellini$^{2}$, 
  M. Colpi$^{1}$, 
  L. Maraschi$^{3}$,\newauthor
  F. Tavecchio$^{2}$,
  R. Decarli$^{4}$, 
  G. Tagliaferri$^{2}$\\
  $^{1}$Universit\`a di Milano - Bicocca, Dip. di Fisica G. Occhialini, Piazza della Scienza 3, I-20126 Milano, Italy\\
  $^{2}$INAF Osservatorio Astronomico di Brera, Via E. Bianchi 46, I-23807 Merate (LC), Italy\\ 
  $^{3}$INAF Osservatorio Astronomico di Brera, Via Brera 28, I-20121 Milano, Italy\\
  $^{4}$Max-Planck-Institut f\"ur Astronomie, K\"onigstuhl 17, D-69117 Heidelberg, Germany\\
}

\begin{document}
\date{Accepted 2011 January 06. Received 2011 January 06; in original
  form 2010 November 16}

\pagerange{\pageref{firstpage}--\pageref{lastpage}} \pubyear{2011}

\maketitle

\label{firstpage}

\begin{abstract}
The recent detection of $\gamma$-ray emission from four radio-loud
narrow-line Seyfert 1 galaxies suggests that the engine driving the
AGN activity of these objects share some similarities with that of
blazars, namely the presence of a $\gamma$-ray emitting, variable, jet
of plasma closely aligned to the line of sight. In this work we
analyze the $\gamma$-ray light curves of the four radio-loud
narrow-line Seyfert 1 galaxies for which high-energy gamma-ray
emission has been discovered by {\it Fermi}/LAT, in order to study
their variability. We find significant flux variability in all the
sources. This allows us to exclude a starburst origin of the
$\gamma$-ray photons and confirms the presence of a relativistic
jet. Furthermore we estimate the minimum {\it e}-folding variability
timescale (3 -- 30 days) and infer an upper limit for the size of the
emitting region (0.2 -- 2 pc, assuming a relativistic Doppler factor
$\delta=10$ and a jet aperture of $\theta=0.1$ rad).
\end{abstract}


\section{Introduction}
Narrow-line Seyfert 1 (NLS1) galaxies are a class of AGN characterized
by a rather narrow width of the H$\beta$ emission line (FWHM $\la$
2000 km/s), and by the presence of a strong FeII bump, a soft X-ray
excess and flux ratio [OIII]/H$\beta <$3
\citep{1985-Osterbrock-spectra_of_nls1, 2000-Pogge-review_nls1}. These
sources are usually radio-quiet, although, in a few cases, they show a
radio-loudness parameter R (ratio between the 5 GHz and optical B flux
densities, \citealt{1989-Kellermann-def_radio_loudness}) greater than
1000 \citep{2006-Komossa-rlnls1_quasar,
  2008-Yuan-population_rlnls1_with_blazar_prop}. Such sources, dubbed
radio-loud narrow-line Seyfert 1 galaxies (RL-NLS1), were thought to
be inactive in $\gamma$-rays, although several authors speculated the
occurrence of similarities with blazars \citep{2003-Zhou-pmnj0948,
  2006-Komossa-rlnls1_quasar, 2007-Zhou-1h0323,
  2008-Yuan-population_rlnls1_with_blazar_prop,
  2009-Foschini-blazar_nuclei_in_rlnls1, 2010-Gu-CompactRadioNLS1}.
The important discovery of $\gamma$-ray emission from the RL-NLS1
source PMN J0948+0022 \citep{2009-Foschini-discovery_pmnj0948,
  2009-Abdo-discovery_pmnj0948} confirmed these similarities, i.e. the
presence of a jet closely aligned to the line of sight as a source of
Compton up-scattered $\gamma$-ray photons. PMN J0948+0022 ($z$=0.585,
\citealt{refz-0948}) is one of the strongest radio source among the
RL-NLS1; for its fast radio variability, source compactness, inverted
radio spectrum and high brightness temperature, PMN J0948+0022 is one
of the first RL-NLS1 for which the presence of a jet has been
hypothesized \citep{2003-Zhou-pmnj0948, 2006-Doi-VLBIObs0948}. The
detection of $\gamma$-ray emission definitively confirms the presence
of a jet and allows to build a complete SED which closely resemble
that of a typical blazar, with two broad peaks in the far IR and in
the $\gamma$-ray range respectively. By modeling the SED with the
synchrotron and inverse-Compton model
\citep{2009-Ghisellini-Canonical_blazar} it is possible to roughly
estimate important parameters such as the black hole mass, the
Eddington ratio, and the power carried by the jet. The resulting
values, although model-dependent, are usually compatible with
estimates found independently in other studies. In the case of PMN
J0948+0022 the black hole mass turns out to be $\sim $1.5 $\times$
10$^8$ M$_{\sun}$ (compatible with the range of values found by
\citealt{2003-Zhou-pmnj0948} using the empirical relation between the
FWHM of emission lines and the continuum luminosity), and the
Eddington ratio which is $\sim$40\%. A $\gamma$-ray variability of a
factor 2.2 rules out the possibility that the $\gamma$-ray emission is
due to a starburst contribution
\citep{2009-Abdo-mw_monitor_pmnj0948}. Shortly after this discovery,
three more RL-NLS1 had been observed in $\gamma$-rays
\citep{2009-Abdo-rlnls1_newclass}: 1H 0323+342, PKS 1502+036 and PKS
2004-447, thus confirming that at least some RL-NLS1 are $\gamma$-ray
emitting AGN. The SED of these sources has also been modeled with the
synchrotron and inverse-Compton model, providing black hole mass
estimates in agreement with previous studies \citep[see][and
  references therein]{2009-Abdo-rlnls1_newclass}. 1H 0323+342
($z$=0.06, \citealt{refz-0323}) is considered to be a composite
nucleus of a NLS1 (for its optical spectrum properties typical of
NLS1) and a blazar (for its flat radio spectrum, core compactness and
X-rays variability, \citealt{2007-Zhou-1h0323}). It is the only
$\gamma$-ray detected RL-NLS1 for which we have an {\it HST}/WFPC2
optical image of the host galaxy. The host shows a ring-like structure
of 15.6 kpc in diameter and the entire galaxy looks like a one-armed
spiral \citep{2007-Zhou-1h0323}. Another study, based on NOT data,
suggests that 1H 0323+342 may be the remnant of a galaxy merger
\citep{2008-Anton-colour1h0323}. The most striking feature of this
source is the high luminosity of the accretion disc which, according
to SED modeling, is approximately 90\% of Eddington value
\citep{2009-Abdo-rlnls1_newclass}. PKS 1502+036 ($z$=0.41,
\citealt{refz-1502}) is the second most powerful source in terms of
Eddington ratio (80 per cent). It carries a jet power, inferred from
SED modeling, sligthly lower than PMN J0948+002
\citep{2009-Abdo-rlnls1_newclass}. Finally, PKS 2004-447 ($z$=0.24,
\citealt{refz-2004}) is an unusual RL-NLS1. It is a strong radio
emitter, with the highest radio-loudness parameter (R$>$6000) among
the four $\gamma$-ray detected RL-NLS1. PKS 2004-447 has an unusually
weak Fe II complex compared to other NLS1 and a rather steep (although
variable) radio spectral index compared to the other
RL-NLS1. \citet{2001-Oshlack-PKS2004} suggests that it may be well
classified as a low-luminosity compact steep spectrum (CSS) radio
quasar, also considering its low inferred black hole mass
($\sim$10$^6$ M$_{\sun}$) \citep{2009-Abdo-rlnls1_newclass}. Spectral
fit of the SED may require a thermal Comptonization component
\citep{2006-gallo-2004} or an external Compton component
\citep{2009-Abdo-rlnls1_newclass}.

The detection of $\gamma$-rays from the four mentioned RL-NLS1 allows
to identify a new class of $\gamma$-ray emitting AGN. The radio
properties (namely temporal variability, flat spectrum and high
brightness temperature) together with the $\gamma$-ray detection
suggests the presence of a relativistic jet closely aligned to the
line of sight. Low power, mildly relativistic and poorly collimated
radio jets have already been observed in a few spiral galaxies hosting
Seyfert nuclei \citep[e.g.][and references
  therein]{2006-Keel-0313-192}, but in the case of PMN J0948+0022 and
PKS 1502+036 the power carried by the jet, as inferred from SED
modeling, is in the range of quasars, while in 1H 0323+342 and PKS
2004-447 is in the range of BL Lac objects
\citep{2009-Abdo-rlnls1_newclass}. Such powerful jets are observed
only in blazars hosted in elliptical galaxies \citep{2009-Marscher}
with black hole masses in the range 10$^8$ -– 10$^9$ M$_{\sun}$. By
contrast black hole masses in $\gamma$-ray detected RL-NLS1 (10$^6$ -–
10$^8$ M$_{\sun}$) are up to three orders of magnitude smaller than
for blazars, thus suggesting that these sources emit at high Eddington
ratios. Furthermore, it seems that the M$_{\rm BH}$-$\sigma_*$ scaling
relation does not apply to NLS1, mainly due to their small FWHM of
permitted lines. \citet{2008-Decarli} showed that a reconciliation
with the scaling relation is possible if the broad line region is
assumed to have a disc-like geometry \citep[but see][]{2008-Marconi}.

While the $\gamma$-ray variability of PMN J0948+0022 has already been
studied in \citet{2009-Abdo-mw_monitor_pmnj0948} and, with a higher
level of significance, during the outburst occurred in July 2010
\citep{2010-Foschini-outburst}, it has never been studied for the
remaining three sources. Aim of this paper is to analyze the {\it
  Fermi}/LAT light curves of the afore-mentioned RL-NLS1 galaxies in
order to put the $\gamma$-ray variability on a firm basis, and to find
the minimum $\gamma$-ray variability timescale.  Throughout the paper,
we assume a $\Lambda$CDM cosmology with H$_0$ = 71 km s$^{-1}$
Mpc$^{-1}$, $\Omega_{\rm m}$ = 0.27, $\Omega_\Lambda$ = 0.73.

\section{Data analysis}
\label{sec-dataanalysis}
\begin{figure*}
\includegraphics[width=8.8cm]{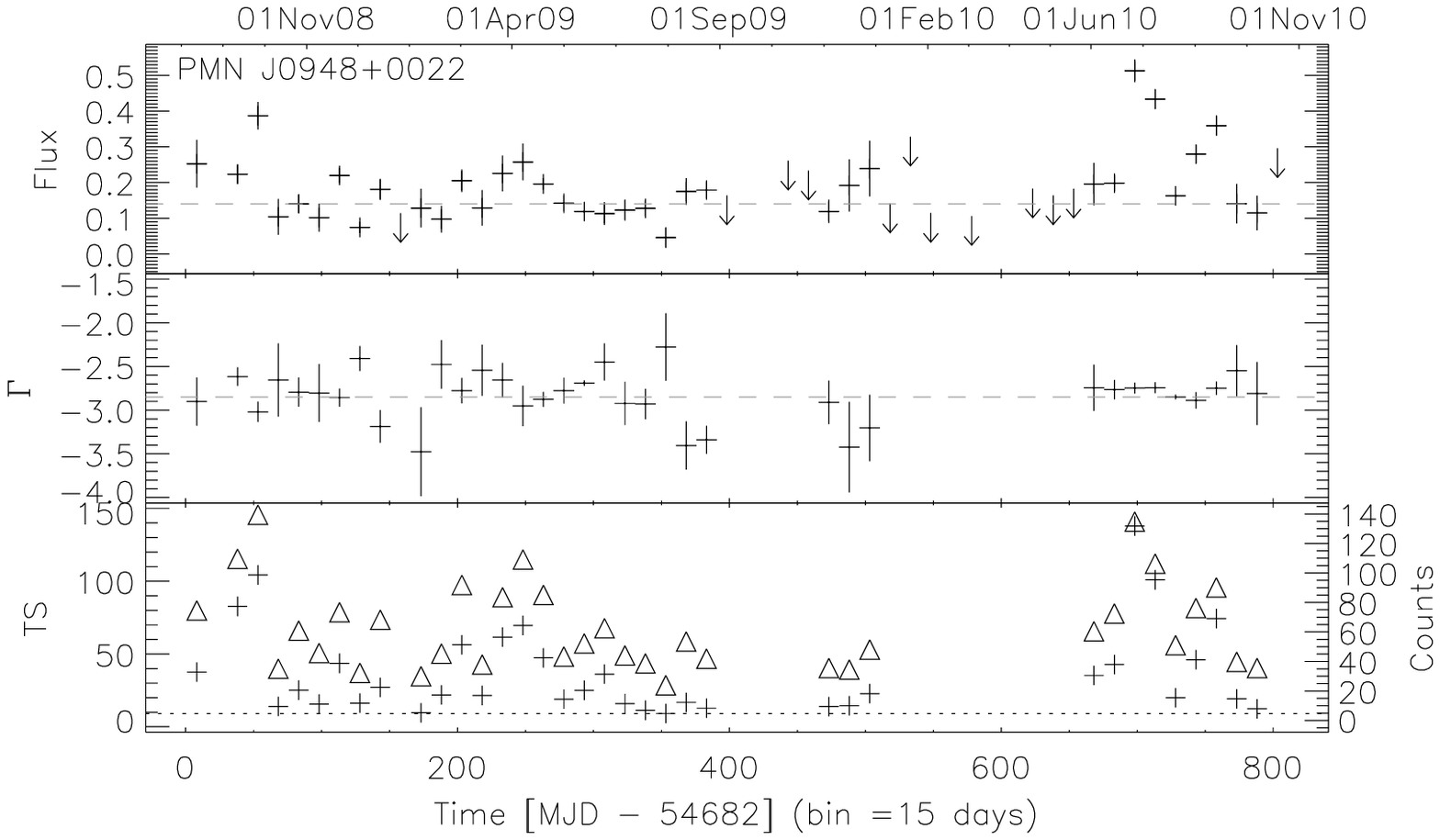}
\includegraphics[width=8.8cm]{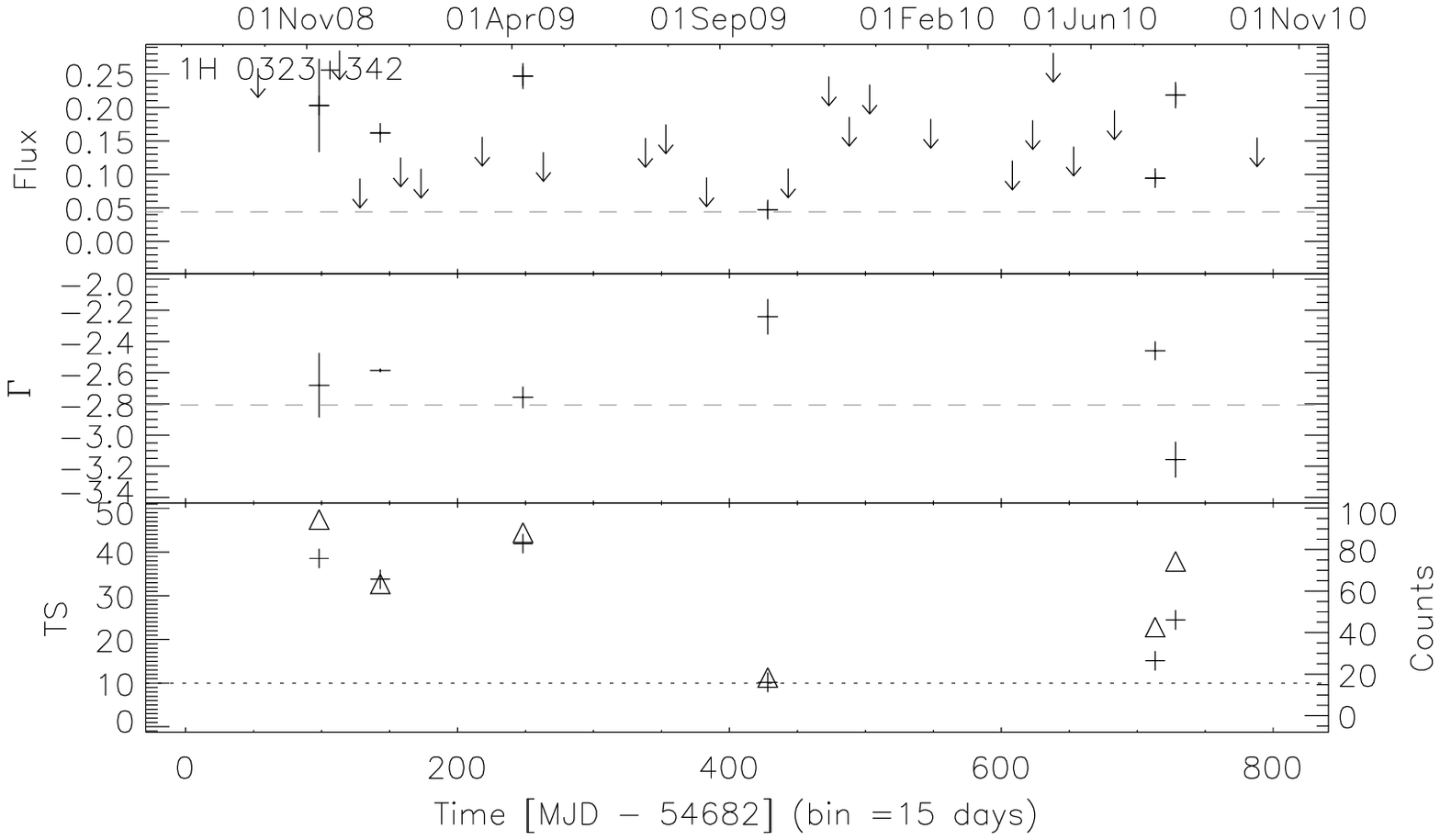}\\
\includegraphics[width=8.8cm]{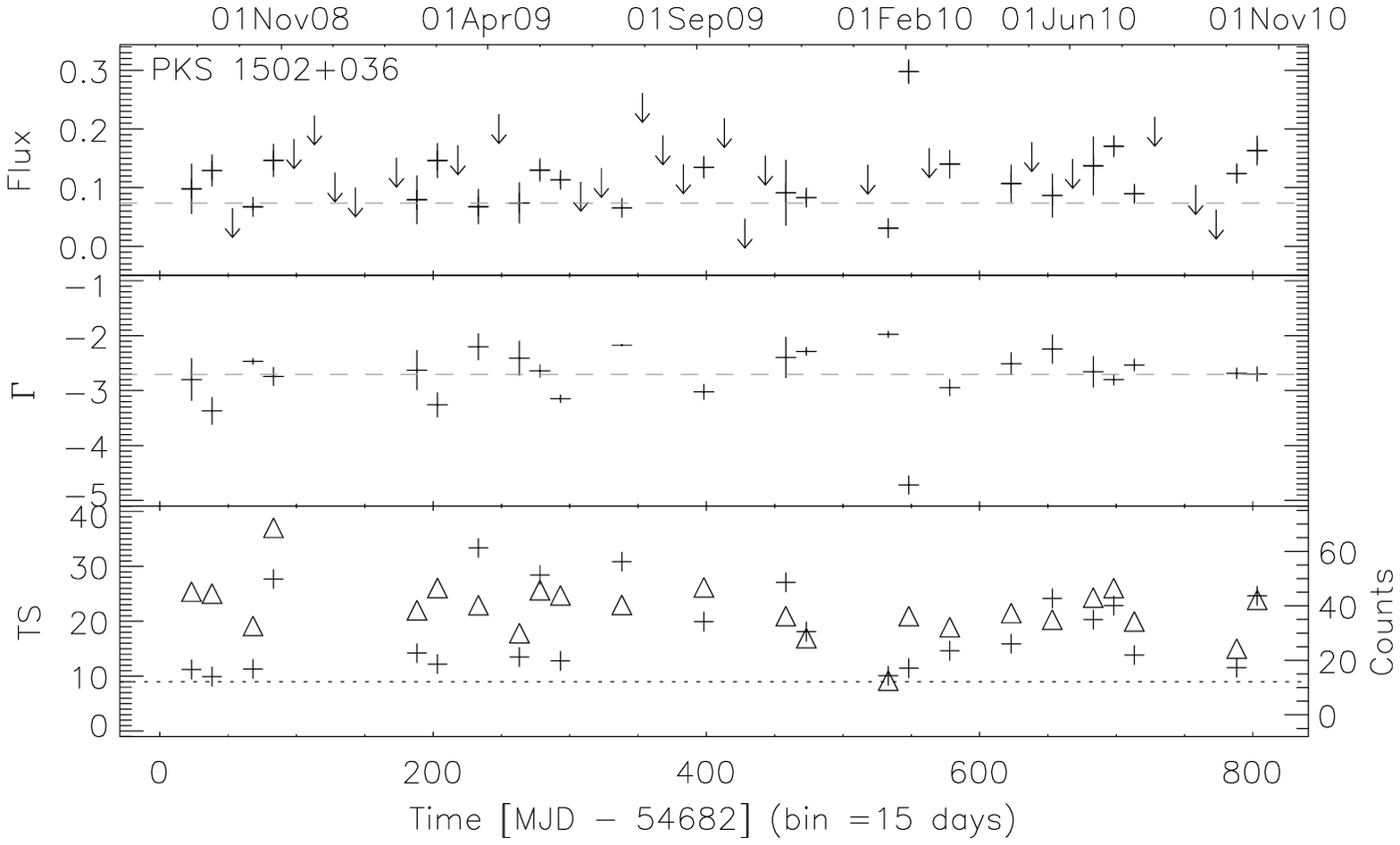}
\includegraphics[width=8.8cm]{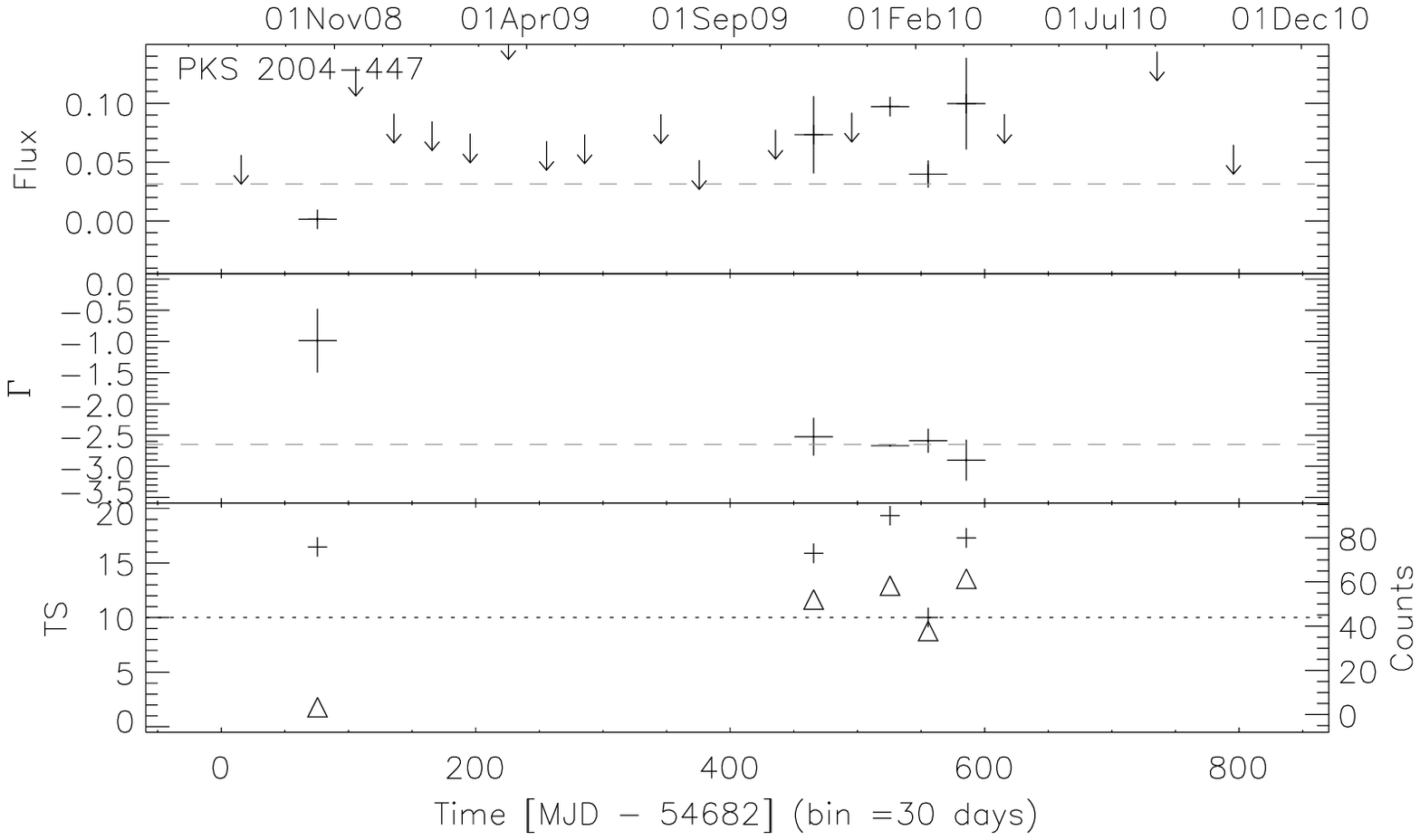}
\caption{Upper panels: light curves of the four RL-NLS1 for detections
  with TS$>$10. Fluxes are given in units of $10^{-6}$ ph cm$^{-2}$
  s$^{-1}$ in the range 0.1 -- 100 GeV. Vertical error bars correspond
  to 1$\sigma$ errors, while horizontal bars corresponds to the time
  binning (15 days for PMN J0948+0022, 1H 0323+342 and PKS 1502+036,
  30 days for PKS 2004-447). Upper limits (TS$<$10) at 2$\sigma$ level
  are denoted by arrows. Middle panels: photon indices. Vertical error
  bars correspond to 1$\sigma$ errors. In both panels horizontal
  dashed lines are the integrated (over the entire period of 26
  months) flux value and photon index respectively. Lower panels: TS
  values (plus symbols) and number of counts (triangle symbols, values
  on the right axis), the horizontal dotted line is the threshold (TS
  = 10).}
\label{fig_lightcurve} 
\end{figure*}

The data analysis has been performed following the standard procedure
described in the {\it Fermi}/LAT documentation. The Science Tools
software version is 9.15.2 and the IRF is \verb|P6_V3_DIFFUSE|. Data
span the period from 2008 august 4 to 2010 october 08 ($\sim$26
months). For each RL-NLS1 all class 3 (diffuse) events coming from a
zenith angle $<105^\circ$ were extracted within an acceptance cone of
radius 10$^\circ$ around the catalog source position. The unbinned
likelihood analysis has been performed modeling the spectrum of each
RL-NLS1 and of all nearby ($<10^\circ$) sources present in the LAT
1-year point source catalog \citep{2010-1FGL} with a power law in the
range 0.1 -- 100 GeV. We computed the integrated flux for each source
analyzing all data over the entire period of 26 months. The extension
of the region of interest to $15^\circ$ around the catalog position
results in non-significant variations of the flux, number of counts
and TS \citep[Test Statistic,][]{1996-Mattox}, as expected since the
PSF of LAT is at most $5^\circ$ wide at 100 MeV. We also performed the
same analysis using alternative spectral models. The use of a broken
power law instead of a simple power law model yields to a slight
increase in the values of TS for PMN J0948+0022 (from 993 to 1194) and
1H 0323+342 (from 86 to 115), and a decrease for PKS 1502+036 and PKS
2004-447. With a log-parabola, the values of TS remain essentially the
same as for the case of a power law for PMN J0948+0022 (from 993 to
999) and PKS 2004-447 (from 97.4 to 98.5), while they decrease for 1H
0323+342 and PKS 1502+036. In the following analysis we will therefore
use the power law models. Then we extracted light curves with time
binning of 15 days (Fig. \ref{fig_lightcurve}; the time binning for
PKS 2004-447 is 30 days) using a TS threshold of 10 (TS$>$10, roughly
equivalent to 3$\sigma$, \citealt{1996-Mattox}) and performed a
chi-squared test against the null hypotesis of constant flux.
If the detection was not significant (TS$<$10) we computed an upper
limit to the flux by varying the source flux value (obtained through
maximization of likelihood) until TS reaches a value of 4
\citep{2010-1FGL}. The resulting fluxes (denoted with arrows in the
figure) corresponds to 2$\sigma$ upper limits. When TS$<$1 we didn't
compute the upper limit since it would be overestimated.

We further proceeded on the analysis of the light curves in order to
compute the minimum {\it e}-folding timescale for each source. We
extracted light curves with different time binnings starting from 30
days. When the detection is significant (TS$>$10) we re-run the
analysis halving the time bin interval, down to a minimum of
approximately 6 hours (roughly corresponding to four {\it Fermi}
orbits, thus ensuring that each source is observed at least twice for
each temporal bin). Then, we considered all combinations of
non-overlapping time bins with the following characteristics: (1) both
time bins have a significant detection, TS$>$10; (2) the flux
difference is greater than the greatest flux error involved at the
3$\sigma$ level; (3) the count difference is significant at the
3$\sigma$ level, assuming a Poisson statistic; (4) the number of
counts in each bin must be greater than 3. For such pairs of bins we
computed the {\it e}-folding timescale as:
\begin{equation}
  \label{eq-tau}
  \tau_{ij} = \left| \frac{t_i - t_j}{\ln{F_i / F_j}} \right|
\end{equation}
where $i$ and $j$ are the indices of the involved time bins
($i>j$). The associated error (at 3$\sigma$ level) is computed through
error propagation:
\begin{equation}
  \label{eq-dtau}
  \Delta \tau_{ij} = \frac{\tau_{ij}}{t_i - t_j}
  \spar{
    \cpar{\Delta t_{ij}}^2 + 
    \cpar{\tau_{ij} \frac{\Delta F_i}{F_i}}^2 + 
    \cpar{\tau_{ij} \frac{\Delta F_j}{F_j}}^2
  }^{1/2}
\end{equation}
where $\Delta t_{ij}$ is half the width of the wider time bin
involved, $\Delta F_i$ and $\Delta F_j$ are the 3$\sigma$ error on
fluxes. We compute the minimum {\it e}-folding variability timescale
as $\tau = \min{\cpar{\tau_{ij}}}$, and associate the corresponding
error $\Delta \tau$ using Eq. \ref{eq-dtau}. The reliability of
Eq. \ref{eq-dtau} has been assessed by simulating 4 series
(respectively for $t_i$, $F_i$, $t_j$ and $F_j$) of 10000 normally
distributed values. The resulting values of $\tau$ are normally
distributed and the 3$\sigma$ confidence intervals are correctly
estimated by $\Delta \tau$ when the mean and standard deviation of
the simulated data are used in Eq. \ref{eq-dtau}.

\begin{figure*}
\includegraphics[width=8cm]{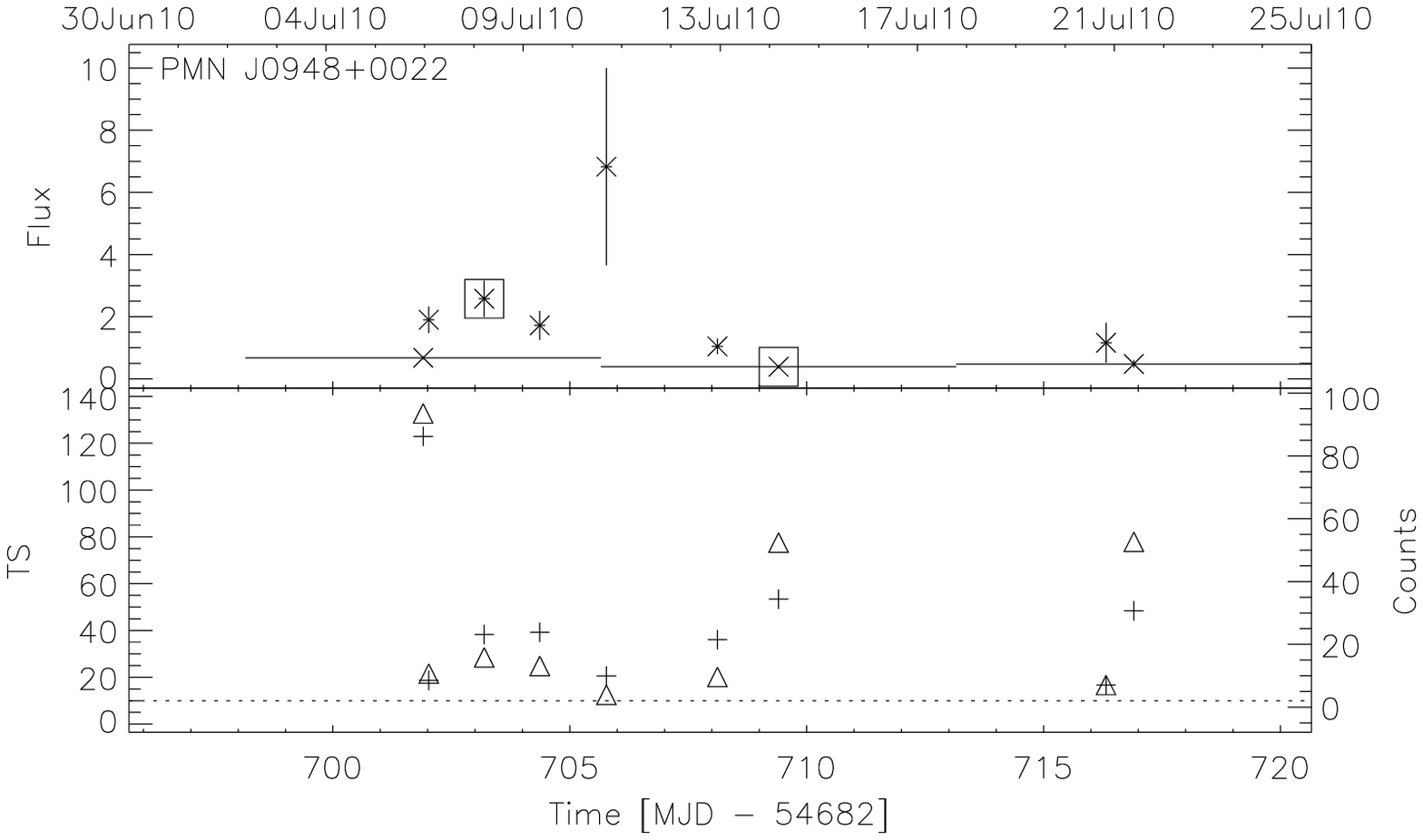}
\includegraphics[width=8cm]{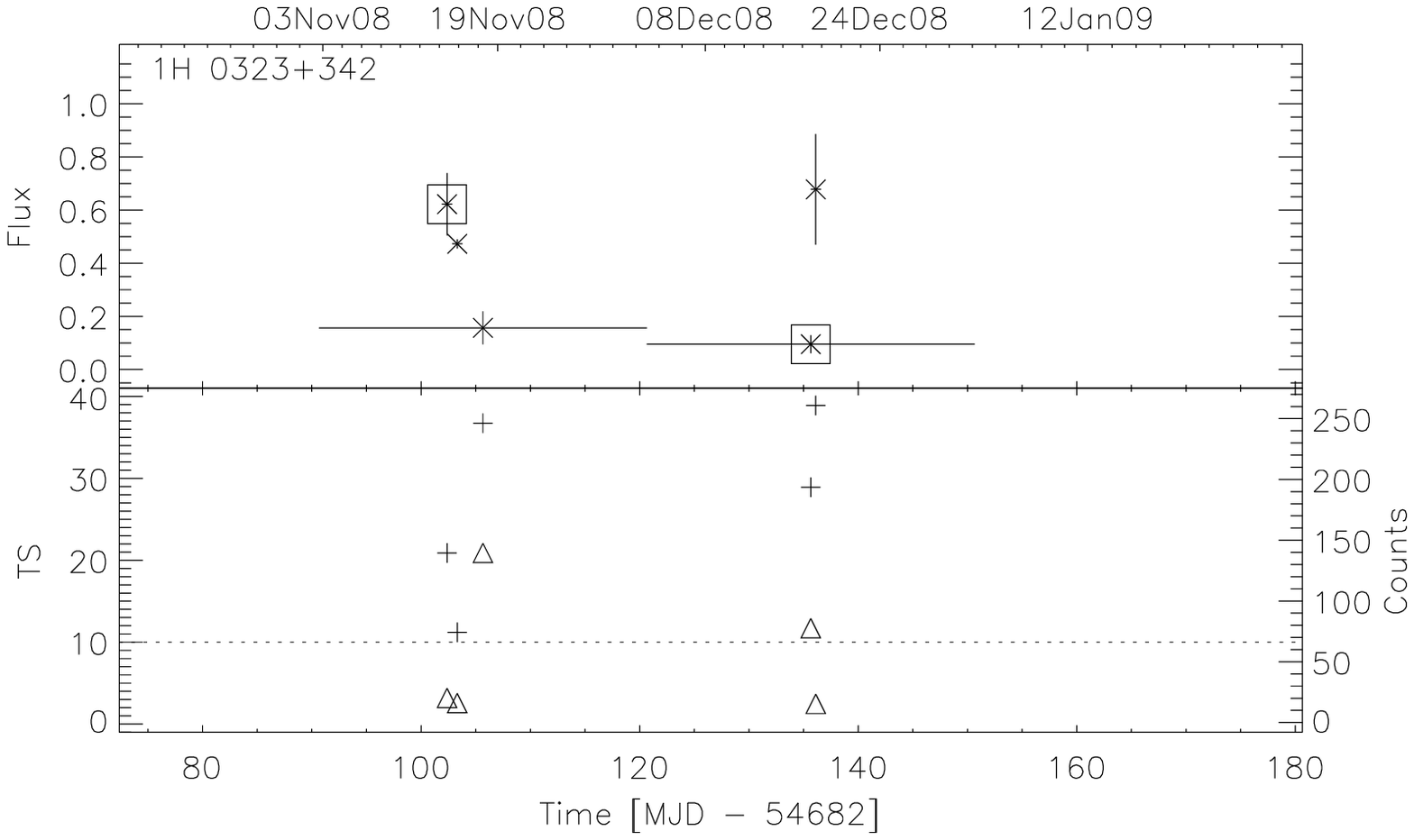}\\
\includegraphics[width=8cm]{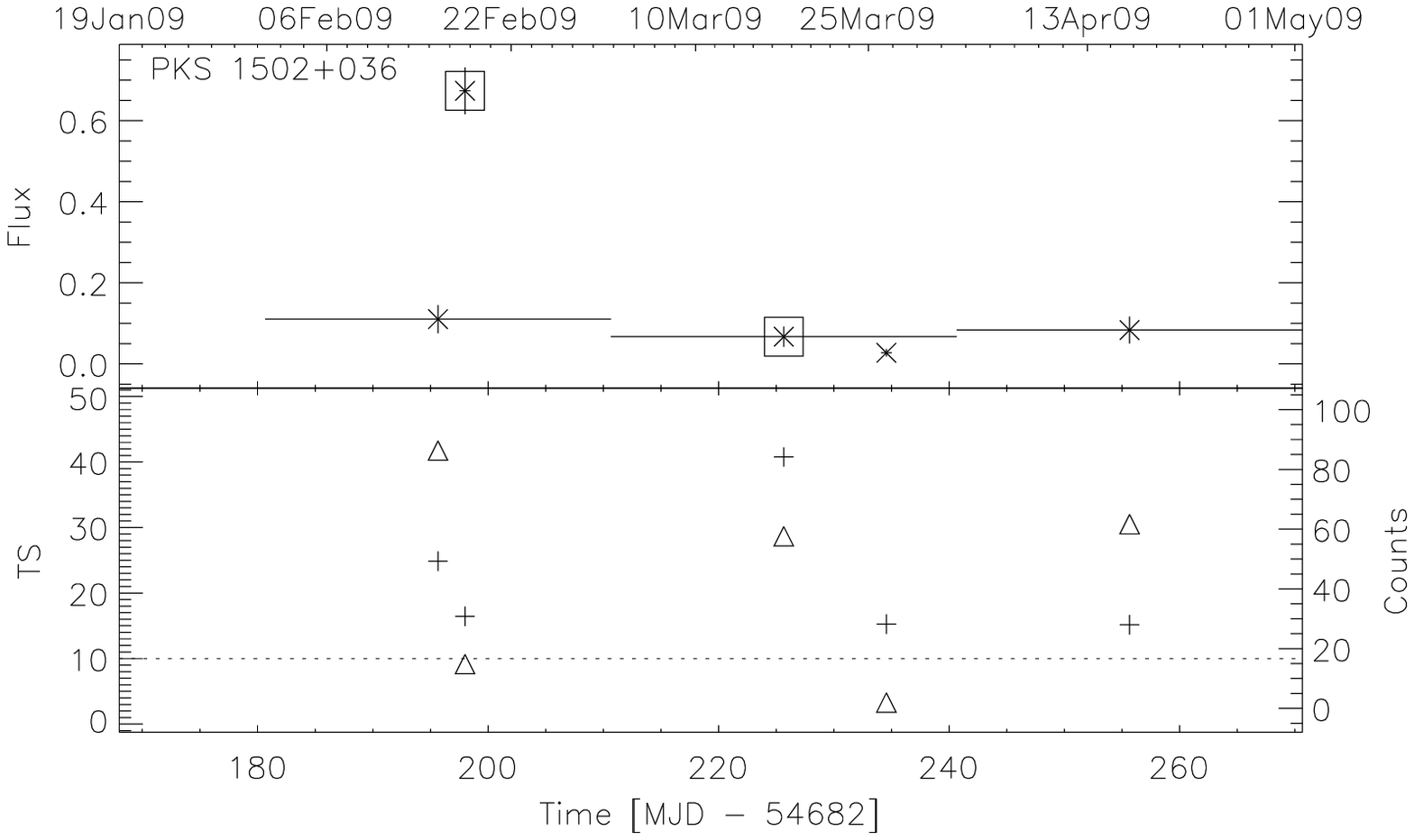}
\includegraphics[width=8cm]{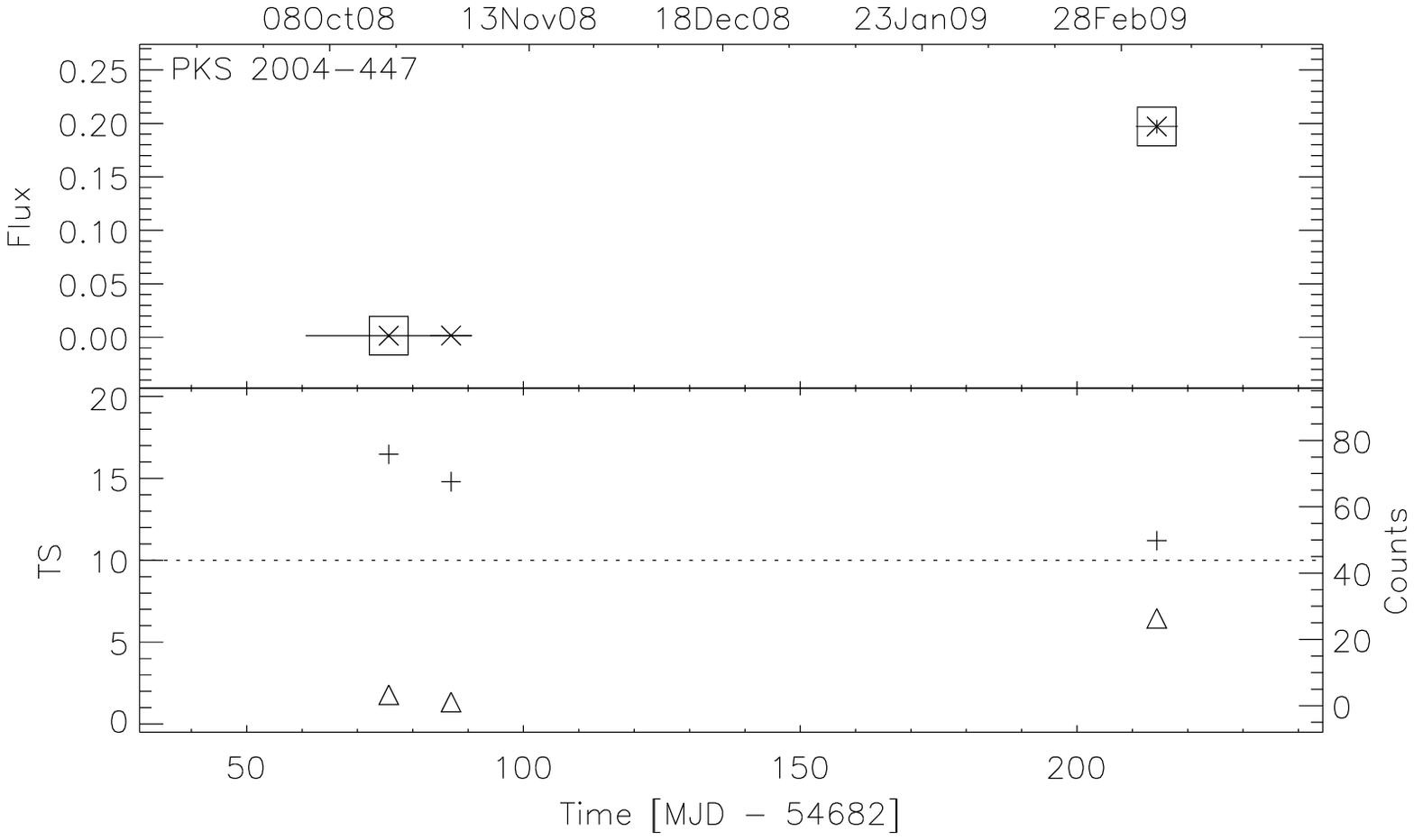}
\caption{Detail view on the bins (denoted by squares) involved in the
  computation of the {\it e}-folding minimum variability
  timescale. Units and meaning of symbols are the same as in
  Fig. \ref{fig_lightcurve} (upper and lower panels). Flux symbols
  have been changed to $\times$ for a clearer visibility. See also
  Table \ref{tab-bins}.}
\label{fig_lightcurvezoom} 
\end{figure*}
\begin{table*}
 \centering
 \begin{minipage}{140mm}
   \caption{Data and results of the analysis on the four RL-NLS1
     sources. Columns are: (1) name of the source; (2) redshift; (3)
     luminosity distance; (4) radio loudness; (5) integrated
     $\gamma$-ray luminosity (0.1 -- 100 GeV) over the entire period
     (26 months) with errors at 1$\sigma$ level; (6) photon index with
     errors at 1$\sigma$ level; (7) $\chi^2$ and (8) DOF computed on
     the light curves of Fig. \ref{fig_lightcurve} in the null
     hypotesis of constant flux equal to the integrated flux; (9)
     minimum {\it e}-folding variability timescale with error at
     3$\sigma$ level.}
   %               12      3      4       5      6        7      8      9        0       1      2       3      4        5
   \label{tab-data}
   \begin{tabular}{lR{1}{2}R{1}{2}R{4}{0}|R{3}{2}@{ }c@{ }R{2}{2}R{3}{3}@{ }c@{ }R{1}{3}|R{5}{0}R{3}{0}|R{3}{1}@{ }c@{ }R{2}{1}}
     \hline
     \multicolumn{1}{c}{Source}                    &
     \multicolumn{1}{c}{z}                         &
     \multicolumn{1}{c}{D$\rm_L$}                  &
     \multicolumn{1}{c}{R}                         &
     \multicolumn{3}{c}{L$_{\gamma}$}              &
     \multicolumn{3}{c}{$\Gamma$}                  &
     \multicolumn{1}{c}{$\chi^2$}                  &
     \multicolumn{1}{c}{DOF}                       &
     \multicolumn{3}{c}{$\tau$}                    \\
     \multicolumn{1}{c}{}                          &
     \multicolumn{1}{c}{}                          &
     \multicolumn{1}{c}{[Gpc]}                     &
     \multicolumn{1}{c}{}                          &
     \multicolumn{3}{c}{[10$^{45}$ erg s$^{-1}$]}  &
     \multicolumn{3}{c}{}                          &
     \multicolumn{1}{c}{}                          &
     \multicolumn{1}{c}{}                          &
     \multicolumn{3}{c}{[days]}                    \\
     \hline
  PMN J0948+0022 &    0.5850 &     3.400 &     1000. &     250.0 &$\pm$&       13.04 &    -2.851 &$\pm$&      0.007130 &     527.6 &   33 &    3.284 &$\pm$&     2.482\\
     1H 0323+342 &   0.06100 &    0.2650 &     151.0 &    0.1563 &$\pm$&     0.03677 &    -2.807 &$\pm$&       0.01004 &     2575. &    5 &    17.72 &$\pm$&     14.40\\
my    PKS 1502+036 &    0.4090 &     2.200 &     1549. &     41.45 &$\pm$&       4.104 &    -2.708 &$\pm$&      0.007048 &     384.9 &   21 &    12.00 &$\pm$&     8.999\\
    PKS 2004-447 &    0.2400 &     1.200 &     6320. &     3.853 &$\pm$&      0.7033 &    -2.650 &$\pm$&      0.005605 &     4032. &    4 &    28.37 &$\pm$&     18.47\\
    \hline
\end{tabular}
\end{minipage}
\end{table*}
\begin{table*}
 \centering
 \begin{minipage}{140mm}
   \caption{Quantities involved in the computation of the minimum {\it
       e}-folding variability timescale.}
   %               12      3      4      5       6      7      8
   \label{tab-bins}
   \begin{tabular}{lR{2}{2}R{2}{2}R{4}{3}@{ }c@{}R{-1}{3}R{2}{1}R{2}{0}}
     \hline
     \multicolumn{1}{c}{Source}                                          &
     \multicolumn{1}{c}{$t$\footnote{MJD - 54682}}                       &
     \multicolumn{1}{c}{$\Delta t$\footnote{Time binning.}}              &
     \multicolumn{3}{c}{F$_{\gamma}$\footnote{Flux in the range 0.1 -- 100 GeV.}}   &
     \multicolumn{1}{c}{N / $\Delta t$}            &
     \multicolumn{1}{c}{TS}                        \\
     \multicolumn{1}{c}{}                          &
     \multicolumn{1}{c}{[days]}                    &
     \multicolumn{1}{c}{[days]}                    &
     \multicolumn{3}{c}{[$10^{-6}$ ph cm$^{-2}$ s$^{-1}$]} &
     \multicolumn{1}{c}{[cts days$^{-1}$]}         &
     \multicolumn{1}{c}{}                          \\
     \hline
  PMN J0948+0022 &     703.2 &    0.2344 &     2.578 &$\pm$&     0.5842 &     67.52 &     38.30\\
                 &     709.4 &     7.500 &    0.3890 &$\pm$&    0.05839 &     6.988 &     53.36\\
  1H 0323+342    &     102.4 &    0.9376 &    0.6221 &$\pm$&     0.1171 &     21.52 &     20.89\\
                 &     135.7 &     30.00 &   0.09514 &$\pm$&    0.03373 &     2.585 &     28.92\\
  PKS 1502+036   &     198.0 &     0.9374 &   0.6737 &$\pm$&    0.05780 &     15.79 &     16.44\\
                 &     225.7 &      30.00 &  0.06717 &$\pm$&    0.02472 &     1.920 &     40.77\\
  PKS 2004-447   &     75.66 &     30.00 &  0.001481 &$\pm$&   0.001470 &    0.1104 &     16.47\\
                 &     214.4 &     7.500 &    0.1971 &$\pm$&   0.006238 &     3.514 &     11.20\\
  \hline
   \end{tabular}
 \end{minipage}
\end{table*}

\section{Results}
\label{sec-results}
Fig. \ref{fig_lightcurve} shows the light curves (upper panels) of the
four RL-NLS1 over the entire period of 26 months. Vertical error bars
are 1$\sigma$ errors on fluxes, while horizontal bars are equal to the
time binning (15 days, except PKS 2004-447 for which is 30 days) for
all points with a significant detection (TS$>$10). Points without a
significant detection are upper limits denoted by arrows. Middle
panels show the photon indices for points with significant detection,
assuming a simple power law model ($F(E) \propto E^\Gamma$) for each
source in the range 0.1 -- 100 GeV. Horizontal dashed lines in both
upper and middle panels are the integrated (over 26 months) flux and
photon index respectively. Lower panels show the TS value (plus
symbol) and number of counts (triangle symbol) for each bin with
significant detection. The integrated luminosity and photon indices,
as well as the $\chi^2$ values and DOF in the hypotesis of constant
flux, are reported in Table \ref{tab-data}. The minimum {\it
  e}-folding variability timescale computed with Eq. \ref{eq-tau} and
\ref{eq-dtau} is reported in Table \ref{tab-data}. The detailed view
(zoom) on the time bins involved in the computation of the minimum
    {\it e}-folding variability timescale are shown in
    Fig. \ref{fig_lightcurvezoom}. To compute $\tau$, points with
    different time binnings are considered and are denoted with
    different horizontal bar lengths in Fig. \ref{fig_lightcurvezoom}
    (e.g. for PMN J0948+0022, one time binning is 0.23 days and the
    other is 7.5 days. The bar corresponding to the shorter time bin
    is barely visible). Similarly, vertical bars refer to 1$\sigma$
    errors in the flux. The square symbols identify the points that
    fulfill the constraints described in Sect. \ref{sec-dataanalysis}
    and provide the value of $\tau$.  Notice, as an example, that the
    point at time $\sim$706 (MJD - 54682, PMN J0948+0022) is excluded
    from the computation since its flux value is compatible with
    nearby points at 3$\sigma$ level (rule (2) of
    Sect. \ref{sec-dataanalysis}). The quantities involved in the
    calculation of $\tau$ (time, time binning, photon flux, number of
    counts and TS values) are reported in Table \ref{tab-bins}.

\section{Discussion}
A statistically significant variability over the entire period of 26
months is present for all sources. The significance of this
variability on timescales of $\la$2 years is supported by the
chi-squared test performed against the null hypotesis of constant flux
(last two columns of Table \ref{tab-data}). This rules out the
possibility that the $\gamma$-ray emission is due to a starburst
activity. Thus, the data support the hypotesis that $\gamma$-ray
photons are associated to the presence of a jet. We cannot exclude the
possibility of a starburst activity but its contribution would be
negligible compared to the jet emission, since the $\gamma$-ray
luminosities (Table \ref{tab-data}) found in our RL-NLS1 are at least
four order of magnitude greater than the archetypal starburst galaxy
M82 hosting a quiescent black hole \citep{1993-Gaffney-mbhM82} and
whose $\gamma$-ray luminosity in the 0.1 -- 100 GeV range is
$\sim$10$^{40}$ erg s$^{-1}$ \citep{2010-Abdo-StarburstM82}.

The next step in our analysis has been to estimate the minimum
timescale variability for each source. Among the many methods to
measure a timescale variability we chose the {\it e}-folding timescale
(Eq. \ref{eq-tau}). The main advantage of this method is that it
allows a computation of the timescale using just two flux
measurements; it does not require any fitting or minimising
procedure. Furthermore the resulting timescale is well defined and can
be used to compare different sources. Finally, the error on the
timescale is easily computed using analytical error propagation
(Eq. \ref{eq-dtau}). The underlying assumption is that the flux evolve
according to an (either increasing or decreasing) exponential
law. Although this assumption is not always justified (e.g. in the
presence of flaring episodes) it is of common practice since we often
do not know the actual law which drives the evolution of the
flux. Furthermore, the requirement of just two flux measurements can
be easily accomplished even for weak sources such as the ones analyzed
in this work. The rules to select the flux measurements used in the
computation of the {\it e}-folding timescale have been described in
Sect. \ref{sec-dataanalysis}: rules (2) and (3) are needed to ensure
that the flux in the two temporal bins are actually different,
otherwise our assumption of exponentially varying flux cannot be
justified. Rules (1) and (4) ensure that the flux measures and
associated errors are reliable. The relative error $\Delta \tau /
\tau$ of our measure of the minimum {\it e}-folding timescale depends
on several factors. The main sources of uncertainties are the errors
in the flux measurements $\Delta F$ and the width of the time bins
$\Delta t$. The ratio $\Delta F / F$ can be made smaller by increasing
the number of counts detected in a wider time bin (greater $\Delta
t$). Viceversa, a narrower time bin results in a smaller number of
counts and consequently a poor accuracy in the flux measurement. The
best achievable accuracy on $\tau$ (that is the smaller value of
$\Delta \tau / \tau$) is thus determined by the trade-off between the
narrowness of the time bin and the accuracy of flux
measurements. Notice however that the narrowness of the time bin is
limited by the requirement that the source is detected with high
significance, TS$>10$ (rule (1) in
Sect. \ref{sec-dataanalysis}). Since the shortest width of the time
bins scales linearly with the inverse of the photon flux, we may use
narrower time bins (and detect shorter timescales) on brighter
sources. The accuracy of $\tau$ is improved if we compute the {\it
  e}-folding timescale using non-contiguous time bins, since the
exponential law is better constrained by distant points rather than
closer ones. On the other hand, if the bins are contiguous, the error
$\Delta \tau$ may be significantly greater than $\tau$ itself,
i.e. the measure would be useless. The issue related to the
computation of $\tau$ with non-contiguous time bins is that we are
deliberately ignoring what lies between the two time bins, even if we
have a significant measure. An example is given in
Fig. \ref{fig_lightcurvezoom} for PMN J0948+0022 (upper-left panel)
where the minimum {\it e}-folding timescale is computed ignoring the
flux measurement at time $\sim$ 706 (MJD - 54682). The reason is that
the flux measure has a rather big error bar, which does not allow us
to consider it statistically incompatible at the 3$\sigma$ level with
neighbouring flux measures (rule (2) in
Sect. \ref{sec-dataanalysis}). This example suggests that the
intrinsic variability timescale, i.e. a measure of ``how fast'' the
source is able to change its flux significantly, may be shorter than
our measure of $\tau$, although we cannot assess it on a firm
statistical basis. This is a consequence of the fact that we are
assuming a particular function (the exponential) and that we are using
only two flux measurements instead of all available data. If our
detector were able to monitor the flux evolution continuously, we
could have employed more sophisticated methods, such as the Fourier
analysis, in order to estimate the intrinsic minimum timescale
variability. Actually, our light curves do not allow a continuous
monitoring of the sources with time binning of 30 days, and we are
thus forced to rely on the {\it e}-folding timescale. As a
consequence, our measure of $\tau$ should be considered as an upper
limit for the intrinsic minimum timescale variability. The sources
display an {\it e}-folding minimum timescale variability in the range
of 3 -- 30 days. In particular, variability of the order of $\sim$days
has been found for PMN J0948+0022, and of the order of $\sim$tens of
days for 1H 0323+342, PKS 1502+036 and PKS 2004-447. The minimum
measured variability timescales in blazars can be as low as 200
seconds at very high energies (E$>$200 GeV) and 800 seconds at X-rays
\citep[see][and references therein]{2007-Aharonian-PKS2155}. At {\it
  Fermi}/LAT energies variability on scales of few hours has been
detected in several blazars
\citep[e.g.][]{2008-ATEL-Foschini-AO_0235+164,
  2010-Tavecchio-variability, 2010-Foschini-3C454.3variability}.  In
particular, the bright blazars 3C 454.3 and PKS 1510-089 showed
variability on scales of 3--6 hours with flux $\sim$10$^{-5}$ ph
cm$^{-2}$ s$^{-1}$ \citep{2010-Tavecchio-variability}, that is 1 -- 2
orders of magnitude greater than the flux of the sources analyzed
here. As discussed above, we expect to measure longer timescales on
weaker sources, as a consequence of the longer time integration
required to significantly detect the source. Our minimum timescale
estimates are indeed approximately 1 -- 2 orders of magnitude longer
compared to those of the above-mentioned blazars. Therefore, we cannot
exclude the possibility that variability in RL-NLS1 is as fast as the
one observed in some of the most luminous blazars. The detection of
shorter timescales on our sources is challenging due to their
weakness. Ground-based Cherenkov telescopes may be more suited to
search for shorter timescale variabilities, if the spectrum of RL-NLS1
extend to very high energies. In principle variability can also be
measured using the flux upper limits and the resulting minimum
timescales would be below the results quoted in Table \ref{tab-data}
(i.e. for PMN J0948+0022 we would obtain $\tau\sim 1$ day). Upper
limit of fluxes are however less reliable than direct flux
measurement, since they depends on the contribution of nearby sources
and of diffuse background.

The minimum {\it e}-folding timescale variability allow to estimate an
upper limit for the size of the emitting region: $R_{\rm blob} \la
\delta c \tau / (1 + z) \sim$0.5 -- 6 $\delta_1 \times$ 10$^{17}$ cm,
where $\delta_1 = \delta / 10$ is the relativistic Doppler factor and
$z$ is the redshift. With a jet aperture $\theta_{-1} = \theta / 0.1$
rad the distance at which most of the kinetic energy of the jet is
dissipated and the $\gamma$-rays observed are produced (the so-called
dissipation region), is $R_{\rm diss} \la$ 0.2 -- 2 $\delta_1 /
\theta_{-1}$ pc. The photon indices are rather steep ($\Gamma < -2$,
with the only exception of the point at time $\sim$75 of PKS
2004-447), thus the inverse Compton peak lies below 100 MeV. In the
framework of the blazar sequence \citep{1998-Fossati-unifiedscheme}
the $\gamma$-ray emitting RL-NLS1 are therefore located in the region
relevant to quasars. The photon index shows also some significant
variations, e.g. at time $\sim$750 (MJD - 54682) the photon index of
1H 0323+342 underwent a 6$\sigma$ variation from $\Gamma$ = - (2.46
$\pm 0.06$) to $\Gamma$ = - (3.2 $\pm$ 0.1) in 15 days (TS = 15 and
24, counts = 43 and 74 respectively). The corresponding photon flux is
larger when the spectrum is steeper, so that the overall $\gamma$-ray
luminosity does not change significantly. Another striking case is
given at time $\sim$550 (MJD - 54682) of PKS 1502+036 for which the
photon index shows a 16$\sigma$ variation from $\Gamma$ = -(1.97 $\pm$
0.06) to $\Gamma$ = -(4.7 $\pm$ 0.2) in 15 days, although the TS
values are much lower (10 and 11 respectively). Also in this case the
overall $\gamma$-ray luminosity does not change significantly. This
behaviour is in contradiction with the harder-when-brigther phenomenon
observed in the X-ray spectra of several HBL \citep{hwb1, hwb2} and at
$\gamma$-ray energies of PMN J0948+0022
\citep{2010-Foschini-outburst}. This phenomenon is usually ascribed to
the upshift of the inverse-Compton peak as the source brightens. The
weakness of the sources prevents us from building a time-resolved
detailed spectra and from drawing any conclusion about the eventual
shift towards lower energies of the inverse-Compton peak in 1H
0323+342 and PKS 1502+036.

\section{Conclusions}
In this work, we report the discovery of a statistically significant
variability in the $\gamma$-ray light curves of the four RL-NLS1
detected with {\it Fermi}/LAT, with minimum variability timescales in
the range 3 -- 30 days. This excludes a potential starburst origin of
the $\gamma$-ray emission, and supports the hypothesis of the presence
of a jet closely aligned to the line of sight.  A hint for photon
index variations on timescales $\sim$tens of days is also found in the
data. Variability appears to be a feature common to the four first
$\gamma$-ray detected RL-NLS1. The minimum timescales found in the
{\it Fermi}/LAT energy range, appropriately scaled with the flux, are
comparable to those found in the most luminous blazars. Thus, it is
not possible to exclude variability as fast as that observed in
blazars. This study goes in the same direction of the finding by
\citealt{2010-Foschini-outburst} who reported compelling evidence of
similarities in the SED shape of PMN J0948+0022 (also analyzed here)
and the archetypal blazar 3C 273. We are confident that in the near
future more RL-NLS1 will be identified as $\gamma$-ray emitters among
the many unidentified $\gamma$-ray sources observed by {\it
  Fermi}/LAT.

\bsp

\label{lastpage}


\begin{thebibliography}{}

\bibitem[\protect\citeauthoryear{{Abdo} et~al.,}{{Abdo}
  et~al.}{2009a}]{2009-Abdo-discovery_pmnj0948}
{Abdo} A.~A.,  et~al., 2009a, \apj, 699, 976

\bibitem[\protect\citeauthoryear{{Abdo} et~al.,}{{Abdo}
  et~al.}{2009b}]{2009-Abdo-mw_monitor_pmnj0948}
{Abdo} A.~A.,  et~al., 2009b, \apj, 707, 727

\bibitem[\protect\citeauthoryear{{Abdo} et~al.,}{{Abdo}
  et~al.}{2009c}]{2009-Abdo-rlnls1_newclass}
{Abdo} A.~A.,  et~al., 2009c, \apjl, 707, L142

\bibitem[\protect\citeauthoryear{{Abdo} et~al.,}{{Abdo}
  et~al.}{2010a}]{2010-Abdo-StarburstM82}
{Abdo} A.~A.,  et~al., 2010a, \apjl, 709, L152

\bibitem[\protect\citeauthoryear{{Abdo} et~al.,}{{Abdo}
  et~al.}{2010b}]{2010-1FGL}
{Abdo} A.~A.,  et~al., 2010b, \apjs, 188, 405

\bibitem[\protect\citeauthoryear{{Aharonian} et~al.,}{{Aharonian}
  et~al.}{2007}]{2007-Aharonian-PKS2155}
{Aharonian} F.,  et~al., 2007, \apjl, 664, L71

\bibitem[\protect\citeauthoryear{{Ant{\'o}n}, {Browne} \&
  {March{\~a}}}{{Ant{\'o}n} et~al.}{2008}]{2008-Anton-colour1h0323}
{Ant{\'o}n} S.,  {Browne} I.~W.~A.,    {March{\~a}} M.~J.,  2008, \aap, 490,
  583

\bibitem[\protect\citeauthoryear{{Brinkmann}, {Papadakis}, {Raeth}, {Mimica} \&
  {Haberl}}{{Brinkmann} et~al.}{2005}]{hwb1}
{Brinkmann} W.,  {Papadakis} I.~E.,  {Raeth} C.,  {Mimica} P.,    {Haberl} F.,
  2005, \aap, 443, 397

\bibitem[\protect\citeauthoryear{{Decarli} et~al.,}{{Decarli}
  et~al.}{2008}]{2008-Decarli}
{Decarli} R.,  et~al., 2008, \mnras, 386, L15

\bibitem[\protect\citeauthoryear{{Doi} et~al.,}{{Doi}
  et~al.}{2006}]{2006-Doi-VLBIObs0948}
{Doi} A.,  et~al., 2006, \pasj, 58, 829

\bibitem[\protect\citeauthoryear{{Drinkwater}, {Webster}, {Francis}, {Condon},
  {Ellison}, {Jauncey}, {Lovell}, {Peterson} \& {Savage}}{{Drinkwater}
  et~al.}{1997}]{refz-2004}
{Drinkwater} M.~J.,  {Webster} R.~L.,  {Francis} P.~J.,  {Condon} J.~J.,
  {Ellison} S.~L.,  {Jauncey} D.~L.,  {Lovell} J.,  {Peterson} B.~A.,
  {Savage} A.,  1997, \mnras, 284, 85

\bibitem[\protect\citeauthoryear{{Foschini} et~al.,}{{Foschini}
  et~al.}{2008}]{2008-ATEL-Foschini-AO_0235+164}
  Foschini L., Longo F., Iafrate G. et al., 2008, ATel 1784


\bibitem[\protect\citeauthoryear{{Foschini} et~al.,}{{Foschini}
  et~al.}{2009}]{2009-Foschini-blazar_nuclei_in_rlnls1}
{Foschini} L.,  et~al., 2009, Advances in Space Research, 43, 889

\bibitem[\protect\citeauthoryear{{Foschini} et~al.,}{{Foschini}
  et~al.}{2010a}]{2010-Foschini-3C454.3variability}
{Foschini} L.,  et~al., 2010a, \mnras, 408, 448

\bibitem[\protect\citeauthoryear{{Foschini} et~al.,}{{Foschini}
  et~al.}{2010b}]{2009-Foschini-discovery_pmnj0948}
  Foschini L. for the Fermi/LAT Collaboration, Ghisellini G., Maraschi L.,
  Tavecchio F., Angelakis E., 2010b, in ``Accretion and ejection in AGN: a
  global view'', eds L. Maraschi, G. Ghisellini, R. Della Ceca \& F. Tavecchio,
  Como (Italy), 22-26 June 2009, ASP Conference Series 427, p. 243


\bibitem[\protect\citeauthoryear{{Foschini} et~al.,}{{Foschini}
  et~al.}{2010c}]{2010-Foschini-outburst}
{Foschini} L.,  et~al., 2010c, MNRAS submitted, \texttt{arXiv:1010.4434}

\bibitem[\protect\citeauthoryear{{Fossati} et~al.,}{{Fossati}
  et~al.}{1998}]{1998-Fossati-unifiedscheme}
{Fossati} G.,  et~al., 1998, \mnras, 299, 433

\bibitem[\protect\citeauthoryear{{Gaffney}, {Lester} \& {Telesco}}{{Gaffney}
  et~al.}{1993}]{1993-Gaffney-mbhM82}
{Gaffney} N.~I.,  {Lester} D.~F.,    {Telesco} C.~M.,  1993, \apjl, 407, L57

\bibitem[\protect\citeauthoryear{{Gallo} et~al.,}{{Gallo}
  et~al.}{2006}]{2006-gallo-2004}
{Gallo} L.~C.,  et~al., 2006, \mnras, 370, 245

\bibitem[\protect\citeauthoryear{{Ghisellini} \& {Tavecchio}}{{Ghisellini} \&
  {Tavecchio}}{2009}]{2009-Ghisellini-Canonical_blazar}
{Ghisellini} G.,  {Tavecchio} F.,  2009, \mnras, 397, 985

\bibitem[\protect\citeauthoryear{{Gu} \& {Chen}}{{Gu} \&
  {Chen}}{2010}]{2010-Gu-CompactRadioNLS1}
{Gu} M.,  {Chen} Y.,  2010, \aj, 139, 2612

\bibitem[\protect\citeauthoryear{{Keel} et~al.,}{{Keel}
  et~al.}{2006}]{2006-Keel-0313-192}
{Keel} W.~C.,  et~al., 2006, \aj, 132, 2233

\bibitem[\protect\citeauthoryear{{Kellermann} et~al.,}{{Kellermann}
  et~al.}{1989}]{1989-Kellermann-def_radio_loudness}
{Kellermann} K.,  et~al., 1989, \aj, 98, 1195

\bibitem[\protect\citeauthoryear{{Komossa} et~al.,}{{Komossa}
  et~al.}{2006}]{2006-Komossa-rlnls1_quasar}
{Komossa} S.,  et~al., 2006, \aj, 132, 531

\bibitem[\protect\citeauthoryear{{Marcha}, {Browne}, {Impey} \&
  {Smith}}{{Marcha} et~al.}{1996}]{refz-0323}
{Marcha} M.~J.~M.,  {Browne} I.~W.~A.,  {Impey} C.~D.,    {Smith} P.~S.,  1996,
  \mnras, 281, 425

\bibitem[\protect\citeauthoryear{{Marconi} et~al.,}{{Marconi}
  et~al.}{2008}]{2008-Marconi}
{Marconi} A.,  et~al., 2008, \apj, 678, 693

\bibitem[\protect\citeauthoryear{{Marscher}}{{Marscher}}{2009}]{2009-Marscher}
  Marscher A., 2009, In: ``The Jet Paradigm - From Microquasars to
  Quasars'', ed T. Belloni, Lect. Notes Phys. 794, in press,
  (\texttt{arXiv:0909.2576})


\bibitem[\protect\citeauthoryear{{Mattox} et~al.,}{{Mattox}
  et~al.}{1996}]{1996-Mattox}
{Mattox} J.~R.,  et~al., 1996, \apj, 461, 396

\bibitem[\protect\citeauthoryear{{Oshlack}, {Webster} \& {Whiting}}{{Oshlack}
  et~al.}{2001}]{2001-Oshlack-PKS2004}
{Oshlack} A.~Y.~K.~N.,  {Webster} R.~L.,    {Whiting} M.~T.,  2001, \apj, 558,
  578

\bibitem[\protect\citeauthoryear{{Osterbrock} \& {Pogge}}{{Osterbrock} \&
  {Pogge}}{1985}]{1985-Osterbrock-spectra_of_nls1}
{Osterbrock} D.~E.,  {Pogge} R.~W.,  1985, \apj, 297, 166

\bibitem[\protect\citeauthoryear{{Pogge}}{{Pogge}}{2000}]{2000-Pogge-review_nls1}
{Pogge} R.~W.,  2000, New Astronomy Review, 44, 381

\bibitem[\protect\citeauthoryear{{Sembay}, {Edelson}, {Markowitz}, {Griffiths}
  \& {Turner}}{{Sembay} et~al.}{2002}]{hwb2}
{Sembay} S.,  {Edelson} R.,  {Markowitz} A.,  {Griffiths} R.~G.,    {Turner}
  M.~J.~L.,  2002, \apj, 574, 634

\bibitem[\protect\citeauthoryear{{Snellen}, {McMahon}, {Hook} \&
  {Browne}}{{Snellen} et~al.}{2002}]{refz-1502}
{Snellen} I.~A.~G.,  {McMahon} R.~G.,  {Hook} I.~M.,    {Browne} I.~W.~A.,
  2002, \mnras, 329, 700

\bibitem[\protect\citeauthoryear{{Tavecchio} et~al.,}{{Tavecchio}
  et~al.}{2010}]{2010-Tavecchio-variability}
{Tavecchio} F.,  et~al., 2010, \mnras, 405, L94

\bibitem[\protect\citeauthoryear{{Williams}, {Pogge} \& {Mathur}}{{Williams}
  et~al.}{2002}]{refz-0948}
{Williams} R.~J.,  {Pogge} R.~W.,    {Mathur} S.,  2002, \aj, 124, 3042

\bibitem[\protect\citeauthoryear{{Yuan} et~al.,}{{Yuan}
  et~al.}{2008}]{2008-Yuan-population_rlnls1_with_blazar_prop}
{Yuan} W.,  et~al., 2008, \apj, 685, 801

\bibitem[\protect\citeauthoryear{{Zhou} et~al.,}{{Zhou}
  et~al.}{2003}]{2003-Zhou-pmnj0948}
{Zhou} H.,  et~al., 2003, \apj, 584, 147

\bibitem[\protect\citeauthoryear{{Zhou} et~al.,}{{Zhou}
  et~al.}{2007}]{2007-Zhou-1h0323}
{Zhou} H.,  et~al., 2007, \apjl, 658, L13

\end{thebibliography}
\end{document}